\documentclass[12pt]{article}

\usepackage{amssymb}
\usepackage{amsmath}
\usepackage{array} 
\usepackage{epsfig}

\begin{document}

\title{\bf 5-dimensional Quantum Gravity Effects in Exclusive
Double Diffractive Events}

\author{A.V. Kisselev\thanks{E-mail address: alexandre.kisselev@mail.ihep.ru}, \
V.A. Petrov\thanks{E-mail address: vladimir.petrov@mail.ihep.ru} \
and R.A. Ryutin\thanks{E-mail address: ryutin@th1.ihep.su} \\
\small Institute for High Energy Physics, 142281 Protvino, Russia}

\date{}

\maketitle

\thispagestyle{empty}

\bigskip

\begin{abstract}
The experimentally measurable effects related to extra dimensional
gravity in a RS-type brane world are estimated. Two options of the
RS framework (namely, with small and large curvature) are
considered. It is shown that both can be detected by the joint
experiment of the CMS and TOTEM Collaborations at the LHC.
\end{abstract}

\section{Massive Gravitons and Radion in the RS model}
\label{sec:RS}

There is no doubt that the discovery of particles like Higgs boson
is the fundamental goal, but it will not solve a very important
problem of the hierarchy between the electro-weak ($246$~GeV) and
Planck ($2.4 \cdot 10^{18}$~GeV) scales. Some models have been
proposed recently, which resolve the problem without supersymmetry
but rather with help of space-time with extra dimensions. For
instance, so-called ADD model~\cite{ADD} ``explains'' the large
value of the Planck scale by the large size of compact extra
dimensions. Such theories open way for many new experimental
studies.

In particular, the model of Randall and Sundrum
(RS)~\cite{RS1,RS2} seems to be the most economic introducing only
one extra dimension which has not to be large. It is based on an
exact solution for gravity in a five-dimensional space-time, where
the extra spatial dimension is a ``folded'' circle. Let $\{ z^M \}
= \{ (x^{\mu}, y) \}$, $M = 0,1,2,3,4$, be its coordinates.
Namely, $y$ is the coordinate along the fifth dimension, while $\{
x^{\mu} \}$, $\mu = 0,1,2,3$, are the coordinate in a
four-dimensional space-time. The background metric of the model is
of the form (contribution of the matter energy-momentum tensor is
neglected):
\begin{equation}\label{02}
ds^2 = \gamma_{MN}(z) \, dz^M dz^N = e^{2 \kappa |y|} \, \eta_{\mu
\nu} \, dx^{\mu} dx^{\nu} + dy^2.
\end{equation}
Here $y = r_c \, \theta$ ($-\pi \leqslant \theta \leqslant \pi$),
$r_c$ being the ``radius'' of the extra dimension, and the
parameter $\kappa$ defines the scalar curvature of the
five-dimensional space-time. The points $(x^{\mu}, y)$ and
$(x^{\mu}, -y)$ are identified, and the periodicity condition,
$(x^{\mu}, y) = (x_{\mu}, y + 2 \pi r_c)$, is imposed. In
Eq.~\eqref{02} $\eta_{\mu \nu}$ is the Minkowski metric.

We consider the so-called RS1 model~\cite{RS1} which has two
4-dimensional branes with equal and opposite tension located at
the point $y = 0$ (called the \emph{TeV brane}, or \emph{visible
brane}) and at $y = \pi r_c$ (referred to as the \emph{Plank
brane}).  All SM fields are constrained to the TeV brane, while
the gravity propagates in all five dimensions (bulk).

Since the warp factor $e^{2 \kappa |y|}$ is equal to 1 on the TeV
brane, four-dimensional coordinates $\{ x^{\mu} \}$ are Galilean
and we have a correct determination of the graviton fields on this
brane. For a zero mode sector of the effective theory, one obtains
a relation between the (reduced) Planck mass and the (reduced)
fundamental gravity scale in five dimensions $\bar{M}_5$:
\begin{equation}\label{04}
\bar{M}_{Pl}^2 = \frac{\bar{M}_5^3}{\kappa} \left( e^{2 \pi \kappa
r_c} - 1 \right).
\end{equation}

In the linear approximation, one can parametrize the metric
$g_{MN}$ as
\begin{equation}\label{06}
g_{MN}(z) = \gamma_{MN}(z) + \frac{2}{\bar{M}_5^{3/2}} \,
h_{MN}(z).
\end{equation}
The invariance of the gravitational action under general
coordinate transformations means that the Lagrangian is invariant
under gauge transformations of the field $h_{MN}(z)$ (for details,
see Refs.~\cite{Charmousis:00,Boos:02}). If we impose so-called
unitary gauge~\cite{Boos:02}, we get:
\begin{equation}\label{08}
h_{\mu 4}(x, y) = 0, \qquad  h_{44}(x, y) = \phi (x),
\end{equation}
where $\phi (x)$ is a massless scalar field which depends on
four-dimensional coordinates only. This new degree of freedom,
called \emph{\textbf{radion}}, corresponds to distance
oscillations between the branes.

This massless scalar field would lead to such a change of the
usual gravitational interaction on the visible brane which is
totally excluded by experimental data. However, if the radion
acquires the mass of the order of 100~GeV~\cite{Goldberger:99},
this will not contradict the experimental data, i.e. the radion
could be the lightest massive scalar excitation of the RS model.

The field $h_{\mu \nu}(x, y)$ (with a radion contribution singled
out) is decomposed into a massless mode $h_{\mu \nu}^{(0)}(x)$
(``classical'' graviton) and ``Kaluza-Klein'' (KK) modes $h_{\mu
\nu}^{(n)}(x)$ which describe \emph{\textbf{massive gravitons}}.
The mass spectrum of the KK gravitons on the TeV brane is the
following:
\begin{equation}\label{10}
m_n = x_n \, \kappa, \qquad n=1,2 \ldots,
\end{equation}
where $x_n$ are zeros of the Bessel function $J_1(x)$, with $x_n
\simeq \pi n$ at large n. The interaction Lagrangian on the
visible (TeV) brane looks like
\begin{equation}\label{12}
\mathcal{L}_{int} = - \frac{1}{\bar{M}_{Pl}} \, T^{\mu \nu} \,
h^{(0)}_{\mu \nu} - \frac{1}{\Lambda_{\pi}} \, T^{\mu \nu} \,
\sum_{n=1}^{\infty} h^{(n)}_{\mu \nu} + \frac{1}{\sqrt{3}
\Lambda_{\pi}} \, T^{\mu}_{\mu} \, \phi \ .
\end{equation}
Here $T^{\mu \nu}$ is the energy-momentum tensor of the matter on
the brane, $h^{(n)}_{\mu \nu}$ is the graviton field with the
KK-number $n$ and mass $m_n$ \eqref{10}. The parameter
\begin{equation}\label{14}
\Lambda_{\pi} = \left( \frac{\bar{M}_5^{3}}{\kappa} \right)^{\!
1/2}
\end{equation}
is a physical scale on the TeV brane. As one can see from
\eqref{12}, the radion field is coupled to the trace of the
energy-momentum tensor.

Let us consider two possibilities to satisfy relation \eqref{04}.
One possibility (we will call it \emph{\textbf{``large curvature
option''}}) is to put
\begin{equation}\label{16}
\kappa \simeq \bar{M}_5 \sim 1 \mathrm{\ TeV},
\end{equation}
that corresponds to $\kappa r_c = 11.3$ in Eq.~\eqref{04}. There
is a series of KK massive graviton resonances, with the lightest
one having a mass of order 1 TeV. As for the radion, it is coupled
rather strongly to the SM fields (mainly, to gluons) since
$\Lambda_{\pi} \sim 1$ TeV.

Another possibility (we will call it \emph{\textbf{``small
curvature option''}}~\cite{Giudice:04,Kisselev:05}) is to take
\begin{equation}\label{18}
\kappa \ll \bar{M}_5 \sim 1 \mathrm{\ TeV}.
\end{equation}
In such a case, the mass splitting $\Delta m \simeq \pi \kappa$
can be chosen smaller than the energy resolution of LHC
experiments. For instance, for $\kappa r_c = 9.7$, we get $\pi
\kappa = 50$ MeV, and the mass of the lightest KK excitation is
$m_1 = 60.5$ MeV. This case is not favorable for the production of
the radion, since the coupling is defined by $\Lambda_{\pi} =
(\bar{M}_5/1 \mathrm{TeV})^{3/2} \, 140$ TeV and is two orders of
magnitude smaller than in the previous case.

\section{Mechanism of the Exclusive Double Diffraction}
\label{sec:EDD}

The main goal of our paper is to analyze experimental
possibilities to detect effects of the extra dimensions in
exclusive double diffractive event (EDDE):
\begin{equation}\label{pp_pXp}
p + p \to p \; + \mbox{``gap''} +  X + \mbox{``gap''} +  p \;,
\end{equation}
where $X$ represents a particle strongly coupled to two-gluon
states (say, the radion, SM Higgs, KK graviton, meson containing
heavy quarks or glueball).

Unique advantages of this process are well-known
(see~\cite{EDDE:menu}-\cite{KMR2:azimuthal} and references
therein): a) clear signature of the process; b) possibility to use
the ``missing mass method'', that improves the mass resolution; c)
background is strongly suppressed; d) spin-parity analysis of the
central system can be done. All these properties can be realized
in common CMS/TOTEM detector measurements at LHC~\cite{TDR:TOTEM}.

The exclusive double diffractive process is related to the
dominant amplitude of the exclusive two-gluon production. Driving
mechanism of this processes is the Pomeron.
\begin{figure}[t]
\epsfysize=6cm \epsffile{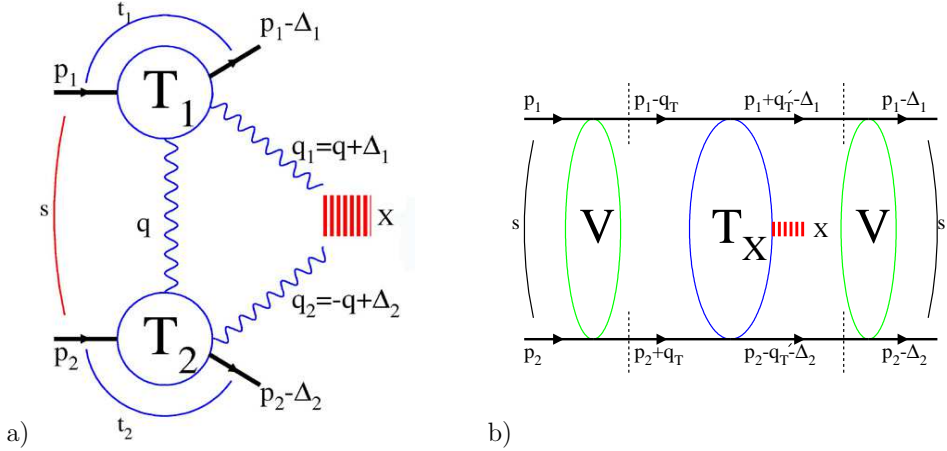} \caption{Model for EDDE: a)
amplitude of the exclusive double diffractive production $T_X$
without unitary corrections; b) amplitude $T_X^{unit}$ with
accounting for ``soft'' re-scattering corrections $V$ in the
initial and final states.} \label{fig:EDDE}
\end{figure}

To calculate an amplitude of the process~\eqref{pp_pXp}, we use an
approach which was considered in detail in Ref.~\cite{EDDE:menu}.
In the framework of this approach, the amplitude can be sketched
as shown in Fig.~\ref{fig:EDDE}. After the tensor contraction of
the amplitudes $T_{1,2}$ with the gluon-gluon fusion vertex, the
full ``bare'' amplitude $T_X$ depicted in Fig.~\ref{fig:EDDE}a
looks like
\begin{equation}\label{Tpp_pXp}
T_X = \frac{2}{\pi} \, c_{gp}^2 \, e^{b(t_1+t_2)}
\left(-\frac{s}{M_X^2}\right)^{\alpha_P(0)} F_{gg \to X} \, I_s \;
.
\end{equation}
Here
\begin{eqnarray}\label{slope}
b &=& \alpha^{\prime}_P(0) \ln \left( \frac{\sqrt{s}}{M_X} \right)
+ b_0 \; ,
\\
b_0 &=& \frac{1}{4} \, (\frac{r^2_{pp}}{2} + r^2_{gp}) \; ,
\end{eqnarray}
with the parameters of the ``hard'' Pomeron trajectory, that
appears to be the most relevant in our case, presented in
Table~\ref{tab:hpomeron}. The last factor in the RHS of
Eq.~\eqref{Tpp_pXp} is
\begin{eqnarray} \label{Isudakov}
I_s &=& \int\limits_{0}^{M_X^2/4}\frac{dl^2}{l^4} \, F_s(l^2)
\left( \frac{l^2}{s_0 + l^2/2} \right)^{2\alpha_P(0)}\; ,
\end{eqnarray}
where $l^2=-q^2\simeq{\mbox{\bf q}}^2$, and $s_0$ is a scale
parameter of the model which is also used in the global fitting of
the data on $pp$ ($p\bar{p}$) scattering for on-shell
amplitudes~\cite{3Pomerons}. The fit gives $s_0 \simeq 1$~GeV$^2$.
If we take into account the emission of virtual "soft" gluons,
while prohibiting the real ones, that could fill rapidity gaps, it
results in a Sudakov-like suppression~\cite{KMR3:sudakov}:
\begin{eqnarray}\label{sudakov}
F_s(l^2) &=& \exp\left[ -\frac{3}{2\pi}
\int\limits_{l^2}^{{M_X}^2/4} \frac{d p_T^2}{p_T^2} \,
\alpha_s({p_T}^2) \ln \left( \frac{{M_X}^2}{4 p_T^2} \right)
\right] \; ,
\end{eqnarray}

The off-shell gluon-proton amplitudes $T_{1,2}$ are obtained in
the extended unitary approach~\cite{Petrov:95}. The ``hard'' part
of the EDDE amplitude, $F_{gg\to X}$, is the usual gluon-gluon
fusion amplitude calculated perturbatively in the SM or in its
extensions.

\begin{table}[h!]
\begin{center}
\caption{Phenomenological parameters of the ``hard'' Pomeron
trajectory obtained from the fitting of the HERA and Tevatron data
(see~\cite{EDDE:menu}, \cite{EVMP:HERA}), and data on $pp$
($p\bar{p}$) scattering~\cite{3Pomerons}.}
\bigskip \bigskip
  \begin{tabular}{||c|c|c|c|c||}
  \hline
  $\alpha_P(0)$ & $\alpha_P^{\prime}(0)$, GeV$^{-2}$ &  $r^2_{pp}$, GeV$^{-2}$
  & $r^2_{gp}$, GeV$^{-2}$ & $c_{gp}$
  \\ \hline
  1.203 &  0.094 &  2.477 &  2.54 & 3.3
  \\ \hline
  \end{tabular}
\label{tab:hpomeron}
\end{center}
\end{table}

The data on total cross-sections demand unambiguously the Pomeron
with larger-than-one intercept, thereof the need in unitarization.
The amplitude with unitary corrections, $T^{unit}_X$, are depicted
in Fig.~\ref{fig:EDDE}b. It is given by the following analytical
expressions:
\begin{eqnarray}\label{ucorr}
T^{unit}_X(p_1, p_2, \Delta_1, \Delta_2) &=& \frac{1}{16\,s
s^{\prime}} \int \frac{d^2\mbox{\bf q}_T}{(2\pi)^2} \,
\frac{d^2\mbox{\bf q}^{\prime}_T}{(2\pi)^2} \; V(s, \mbox{\bf
q}_T) \;
\nonumber \\
&\times& T_X( p_1-q_T, p_2+q_T,\Delta_{1T}, \Delta_{2T}) \,
V(s^{\prime}, \mbox{\bf q}^{\prime}_T) \;,
\\
V(s, \mbox{\bf q}_T) &=& 4s \, (2\pi)^2 \, \delta^2(\mbox{\bf
q}_T) + 4s \!\! \int d^2\mbox{\bf b} \, e^{i\mbox{\bf q}_T
\mbox{\bf b}} \left[e^{i\delta_{pp\to pp}}-1\right]\;,
\end{eqnarray}
where $\Delta_{1T} = \Delta_{1} -q_T - q^{\prime}_T$, $\Delta_{2T}
= \Delta_{2} + q_T + q^{\prime}_T$, and the eikonal function
$\delta_{pp\to pp}$ can be found in Ref.~\cite{3Pomerons}. Left
and right parts of the diagram in Fig.~\ref{fig:EDDE}b denoted by
$V$ represent ``soft'' re-scattering effects in initial and final
states, i.e. multi-Pomeron exchanges. As was shown in
\cite{EDDE:glueballs}, these ``outer'' unitary corrections
strongly reduce the value of the corresponding cross-section and
change an azimuthal angle dependence.

\section{Exclusive Particle Production in Double Diffractive Processes}

\subsection{Radion Production in EDDE}
\label{subsec:radion}

In this subsection, we will study the radion production in EDDE in
the case of the ``large curvature option'' \eqref{16}, taking into
account an effect of a mixing between the radion and SM Higgs.%
\footnote{A similar problem for an inclusive production was
considered in \cite{Battaglia:03}.}

Since the radion has the same quantum numbers as the standard
Higgs boson $h$, there could be the mixing between them which
leads to significant changes in the cross-section of
hadroproduction of both~\cite{Giudice:01}. This problem and its
phenomenological consequences were studied by many authors (see,
for example, \cite{Chaichian:02}).
\begin{figure}[ht]
\epsfysize=6cm \epsffile{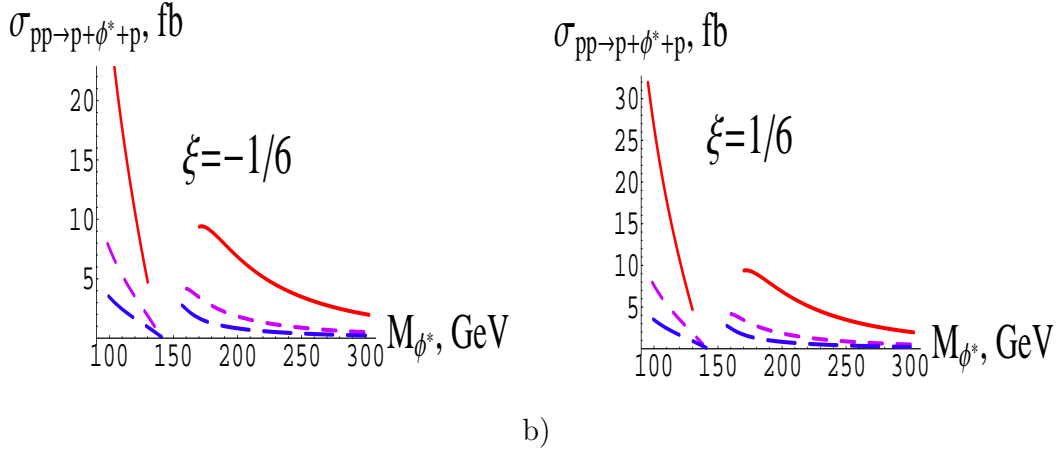}
\caption{The cross section for the production of the radion in
EDDE vs. mass of the observable eigenstate $\phi^*$. Mass
parameter of the Higgs is $M(h)=150$~GeV. Three curves
correspond (from top to bottom) to $\Lambda_{\phi} = $ 1~TeV,
2~TeV, 4~TeV. a) $\xi=-1/6$; b) $\xi=1/6$.}
\label{fig:cstotr}
\end{figure}

The following Lagrangian describes the Higgs-radion interaction:
\begin{equation}\label{mixing}
\mathcal{L}_{h-\phi} = - \frac{6 \, \xi \upsilon}{\Lambda_{\phi}}
\, \phi \, \square \, h.
\end{equation}
Here $\upsilon = 246$ GeV and $\Lambda_{\phi} = \sqrt{3}
\Lambda_{\pi}$ are the vacuum expectation values of the Higgs and
radion fields, respectively. The quantity $\xi$ is the
Higgs-radion mixing parameter. For $\xi = 0$, the radion decouples
from the Higgs. If $\xi \neq 0$, the radion and the SM Higgs boson
mix into the two new eigenstates. Branching fractions for
transitions into SM states can be quite different depending on
$\xi$ and $\Lambda_{\phi}$.

The results of our calculations of the total cross sections are
presented in Fig.~\ref{fig:cstotr}a and \ref{fig:cstotr}b for
various values of the mixing parameter $\xi$ and $\Lambda_{\phi}$.
As we see from these figures, due to the Higgs-radion mixing, the
cross-sections for a single radion production in EDDE can be
larger than those for the SM Higgs boson. Estimated number of
events per year for the integrated luminosity $30$~fb$^{-1}$ and
registration efficiency 10\% are given in Table~\ref{tab:radion}.
\setlength{\extrarowheight}{4pt}

\begin{table}[h!]
\begin{center}
\caption{The expected number of EDDE's for the production of the
radion for various values of $\Lambda_{\phi}$, and $\xi=1/6$
($-1/6$). Higgs mass parameter $M(h)$ is set to 150~GeV. A realistic
value of the total efficiency of an event registration is assumed
to be 10\% (as it was
estimated from fast Monte-Carlo simulation for
the SM Higgs case), and integrated luminosity is taken to be
30~fb$^{-1}$.}
\bigskip \bigskip
  \begin{tabular}{||c||c|c|c||}
  \hline
   & \multicolumn{3}{c||}{$M_{\phi^*}$, GeV}
  \\ \hline
  $\Lambda_{\phi}$, TeV & 100 & 120 & 130
  \\ \hline
  1 & 116 (78)  & 64 (31) & 48 (13)
  \\ \hline
  2 & 31 (22) & 20 (10) & 17 (4)
  \\ \hline
  4 & 8 (5) & 6 (3) & 5 (1)
  \\ \hline
  \end{tabular}
\label{tab:radion}
\end{center}
\end{table}

\begin{figure}[t]
\centering\epsfysize=5cm \epsffile{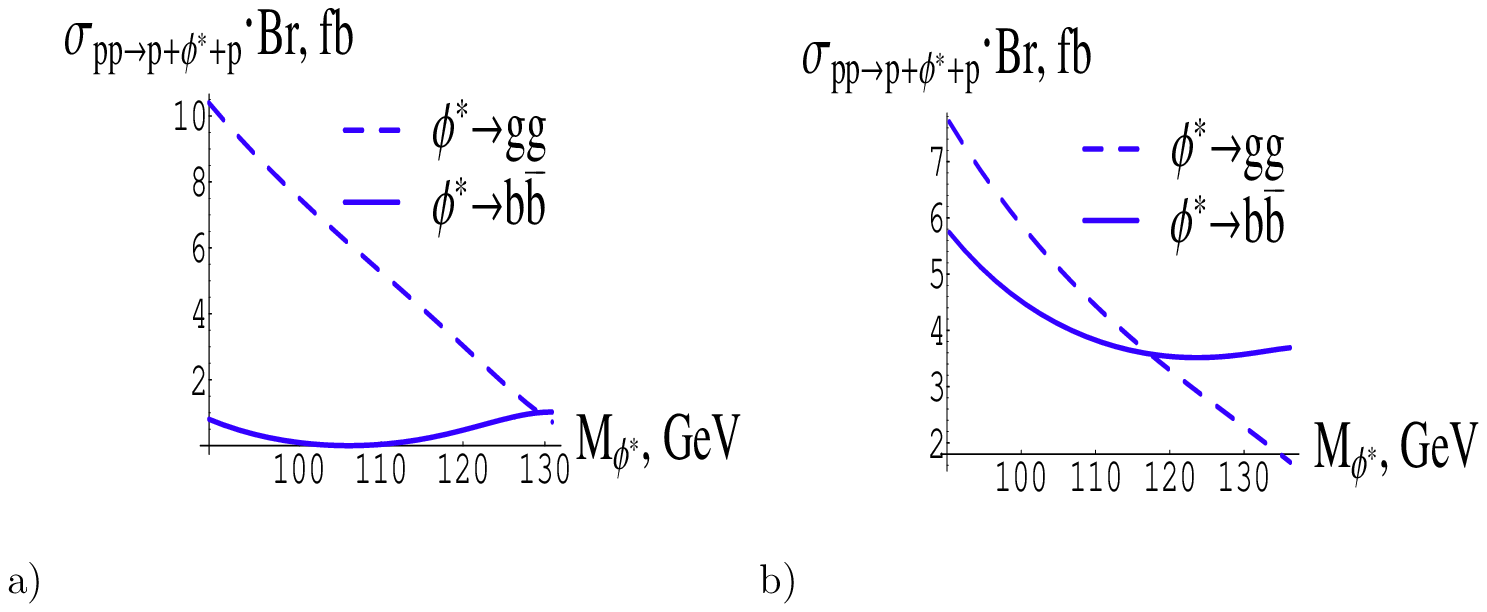}
\caption{The cross section for the production of the radion in EDDE
multiplied by the branching fraction vs. mass of the observable
eigenstate $\phi^*$. The curves correspond to $gg$ decay mode (dash
one) and $b\bar{b}$ (solid one). The parameters of the model are: a)
$\Lambda_{\phi} = $ 2~TeV, $M(h)=$ 150~GeV, $\xi=-1/6$; b)
$\Lambda_{\phi} = $ 2~TeV, $M(h)=$ 150~GeV, $\xi=1/6$.}
\label{fig:csBrsmall}
\vskip 0.2cm
\centering\epsfysize=5cm \epsffile{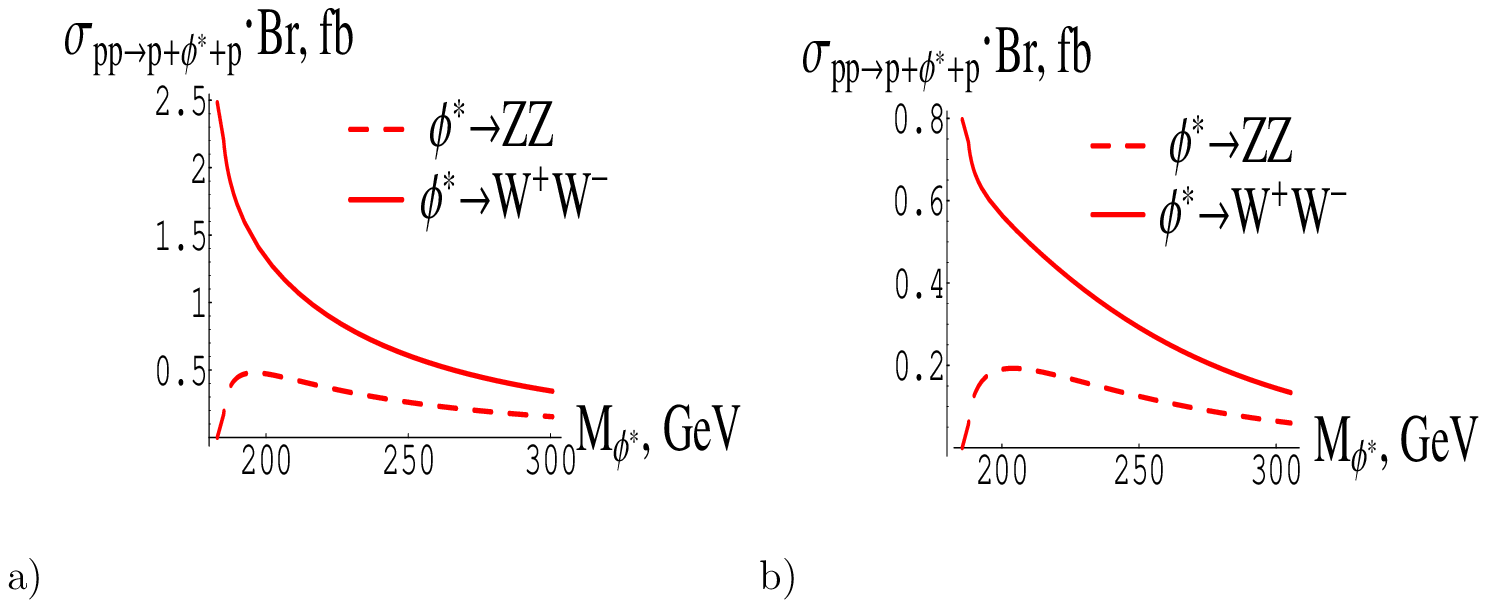}
\caption{The cross section for the production of the radion in EDDE
multiplied by the branching fraction vs. mass of the observable
eigenstate $\phi^*$. The curves correspond to $ZZ$ decay mode (dash
one) and $W^+W^-$ (solid one). The parameters of the model are: a)
$\Lambda_{\phi} = $ 2~TeV, $M(h)=$ 150~GeV, $\xi=-1/6$; b)
$\Lambda_{\phi} = $ 2~TeV, $M(h)=$ 150~GeV, $\xi=1/6$.}
\label{fig:csBrlarge}
\end{figure}

In Figs.~\ref{fig:csBrsmall} and \ref{fig:csBrlarge} one can see
$\sigma (\phi^*)\cdot Br$ for various decay modes. The values of
the parameters are $\xi=1/6,\; -1/6$ and $\Lambda_{\phi}=$ 2~TeV,
$M(h)=$ 150~GeV. A significance of the event is estimated to be
greater than $3\sigma$ in the region of $M_{\phi^*}<$ 140~GeV
($b\bar{b}$ decay mode) and $\Lambda_{\phi}<$ 5~TeV. For larger
radion masses ($ZZ$ and $W^+W^-$ decay modes), we get a similar
significance for the integrated luminosity 100~fb$^{-1}$.

\subsection{Production of the KK Gravitons in EDDE}
\label{subsec:gravitons}

As was shown in Ref.~\cite{Kisselev:05}, the ``small curvature
option'' of the RS model is similar to the ADD model of gravity in
a flat space-time with one compact extra dimension~\cite{ADD} up
to the following formal replacement in the KK sector:
\begin{equation}\label{ADD_RS}
\bar{M}_{Pl} \rightarrow \Lambda_{\pi}, \qquad R_c \rightarrow
\frac{1}{\pi \kappa}.
\end{equation}
Here $R_c$ is a radius of the extra dimension in the flat
space-time.

As was already mention in the end of Section~\ref{sec:RS}, this
case is not favorable for the production of the radion, due to a
large value of the scale $\Lambda_{\pi}$ (remember that it defines
the coupling of the radion to the SM fields, see the last term in
Eq.~\eqref{12}). On the contrary, KK graviton production cross
section does not depend on $\Lambda_{\pi}$, but it is defined by
the 5-dimensional Planck scale $\bar{M}_5$ only, as it will be
demonstrated below.

In the case of a small value of the curvature parameter $\kappa$
\eqref{18}, we have the spectrum of the KK gravitons with the
small mass splitting. Since the widths of the massive gravitons,
$\Gamma_n$, are very small~\cite{Kisselev:04},
\begin{equation}\label{graviton_width}
\frac{\Gamma_n}{m_n} \simeq \left( 0.31 \,
\frac{m_n}{\Lambda_{\pi}}\right)^2,
\end{equation}
the KK gravitons behave like \emph{\textbf{extremely narrow
massive spin-2 resonances}}. Thus, a typical collider signature of
the KK graviton production is \emph{\textbf{an imbalance in
missing mass of final states with a continuous mass
distribution}}, which could be observed in the EDDE-like process
of the type
\begin{equation}\label{pp_p_graviton_p}
p + p \rightarrow p + \mbox{``nothing''} + p \; .
\end{equation}
In other words, one should look for a double diffractive process
with a missing mass $M_{miss}$ and ``nothing else'' in the central
region.

It can be shown that the distribution in $M_{miss}$ is proportional to
\begin{equation}\label{distribution}
\frac{d \sigma_{gr}}{d M_{miss}} \sim \frac{1}{\kappa  \,
\Lambda_{\pi}^2} \sim \frac{1}{\bar{M}_5^3}.
\end{equation}
In other words, both the distribution in $M_{miss}$ and total
cross section $\sigma_{gr}$ are defined by the fundamental Planck
scale in five warped dimensions only, not by the values of
$\kappa$ and $\Lambda_{\pi}$ separately. Since $\bar{M}_5 \sim 1$
TeV, we expect that corresponding cross sections will be large
enough to be measured in the common experiment of the CMS and
TOTEM Collaborations~\cite{TDR:TOTEM}.
\begin{figure}[t!]
\centering \epsfysize=6cm \epsffile{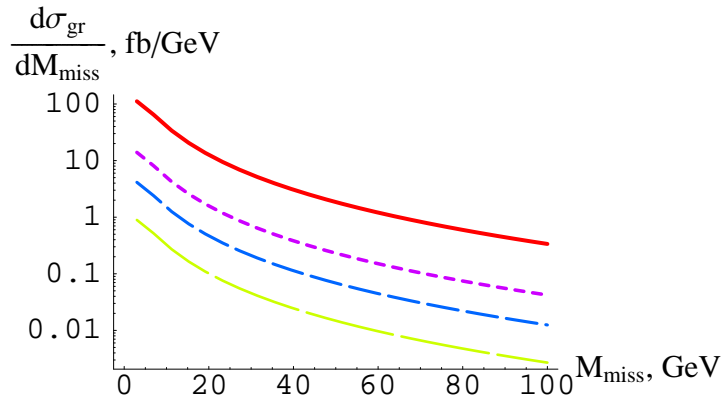} \caption{ The
distribution in the missing mass in the double diffractive
production of the KK gravitons in the RS model with the small
curvature. The curves correspond (from top to bottom) to
$\bar{M}_5 = $ 1~TeV, 2~TeV, 3~TeV, and 5~TeV.}
\label{fig:miss_mass_dist}
\end{figure}

\begin{figure}[ht!]
\vskip 0.2cm \centering \epsfysize=6cm \epsffile{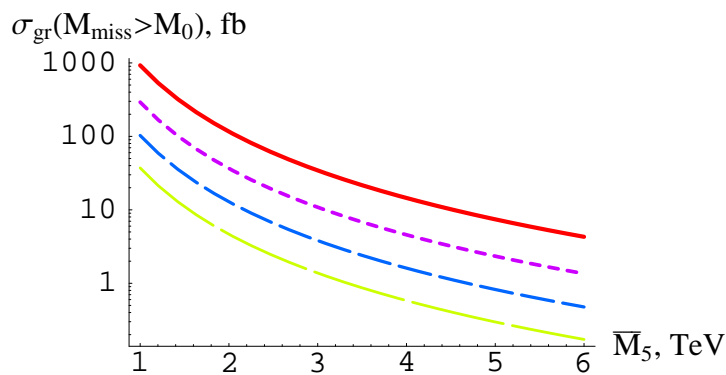}
\caption{The cross section for double diffractive production of
the KK gravitons with masses larger that $M_0$ as a function of
5-dimensional Planck scale $\bar{M}_5$. The curves correspond
(from top to bottom) to $M_0 = $ 3~GeV, 14~GeV, 30~GeV, and
50~GeV.}
\label{fig:grav_cross_sec}
\end{figure}

To estimate $d \sigma_{gr}/d M_{miss}$ and the total cross section
of the graviton production in EDDE, $\sigma_{gr}$, numerically we
use the model of EDDE described in Section \ref{sec:EDD}. The
results of our calculations are presented in
Figs~\ref{fig:miss_mass_dist}, \ref{fig:grav_cross_sec} and
Table~\ref{tab:event_number}. The missing mass distribution $d
\sigma_{gr}/d M_{miss}$ is shown in Fig.~\ref{fig:miss_mass_dist}.
In Fig.~\ref{fig:grav_cross_sec} one can see $\sigma_{gr}$ as a
function of $\bar{M}_5$. The curves in
Fig.~\ref{fig:grav_cross_sec} correspond to various values of
$M_0$, a lower cut on $M_{miss}$, which is imposed to suppress the
soft photon/graviton contributions, and to provide applicability
of our mechanism of exclusive double diffractive production. The
cut of $M_0$ is chosen to be 3, 14 and 30 GeV. An upper cut on
$M_{miss}$ is taken to be 90 GeV in order to suppress a possible
background from the process $p + p \rightarrow p \, + \,
\mbox{neutrinos} \, + \, p$. Note that this upper cut-off does not
significantly reduce a signal due to a rapid fall-off of $d
\sigma_{gr}/d M_{miss}$ in variable $M_{miss}$ (see
Fig.~\ref{fig:miss_mass_dist}).

Let us underline that the results obtained in this Section do not
depend on the physical scale $\Lambda_{\pi}$ (or curvature
parameter $\kappa$), but depend only on the fundamental gravity
scale in five dimensions $\bar{M}_5$.

\begin{table}[t!]
\begin{center}
\caption{The integrated luminosity related to $M_0$, possible
lower experimental limit on missing mass
measurements~\cite{TDR:TOTEM}. Here the total efficiency is
assumed to be 100\%. To get more realistic estimation, one needs
further Monte-Carlo simulations.}
\bigskip \bigskip
  \begin{tabular}{||c||c|c|c|c||}
  \hline
  $M_0$, GeV & 3 & 14 & 30 & 50
  \\ \hline
  $\mathcal{L}$, fb$^{-1}$ & 0.3 & 0.3 - 30 & 30 - 300 & 30 - 300
  \\ \hline
  \end{tabular}
\label{tab:cut_vs_lum}
\end{center}
\end{table}
\begin{table}[t!]
\begin{center}
\caption{The expected number of EDDE's for the production of the
KK gravitons for various values of the fundamental gravity scale
$\bar{M}_5$ and parameter $M_0$. The integrated luminosity is
taken from Table~\ref{tab:cut_vs_lum}.}
\bigskip \bigskip
  \begin{tabular}{||c||c|c|c|c||}
  \hline
   & \multicolumn{4}{c||}{$M_0$, GeV}
  \\ \hline
  $\bar{M}_5$, TeV & 3 & 14 & 30 & 50
  \\ \hline
  1 & 280  & $87-8.7 \cdot 10^3$ &
  $3.0 \cdot 10^3-3.0 \cdot 10^4$ & $1.11 \cdot 10^3-1.11 \cdot 10^4$
  \\ \hline
  2 & 36 & $11-1.1 \cdot 10^2$  &  $390-3.9 \cdot 10^3$
  & $138-1.38 \cdot 10^3$
  \\ \hline
  3 & 9 & $3-300$ &
  $114-1.14 \cdot 10^3$ & $42-420$
  \\ \hline
  5 & 3 & $0.7-70$ &
  $24-240$ & $9-90$
  \\ \hline
  \end{tabular}
\label{tab:event_number}
\end{center}
\end{table}

\section{Conclusions}

In this paper we present the evaluation of experimentally
observable signals due to extra dimensional gravity effects in a
RS-type brane world.

Our estimates show that both extreme options of the RS-type
scenario (i.e. those of small and large curvature) give
distinctive signals which could be fairly detected at the LHC
(possible joint measurements by the CMS and TOTEM LHC
Collaborations).

\section*{Acknowledgements}
This work is supported by grants PICS2910 of the CNRS and
RFBR-04-02-17299.


\end{document}